\documentclass[journal=jctccc,manuscript=article,layout=twocolumn]{achemso}

\usepackage[version=3]{mhchem} 
\usepackage{subfig}

\usepackage{lineno}
\usepackage{multicol,caption,lipsum}

\let\oldmaketitle\maketitle
\let\maketitle\relax

\author{Jayashrita Debnath}
\affiliation[]{Department of Chemistry and Applied Biosciences, ETH Z\"urich, Switzerland}
\alsoaffiliation{Istituto Italiano di Tecnologia, Via Morego 30, 16163 Genova, Italy}
\author{Michele Parrinello}
\affiliation{Istituto Italiano di Tecnologia, Via Morego 30, 16163 Genova, Italy}
\email{michele.parrinello@iit.it}

\title[An \textsf{achemso} demo]
  {Computing rates and understanding unbinding mechanism in host-guest systems}

\abbreviations{}

\begin{document}



\twocolumn[
\begin{@twocolumnfalse}
	\oldmaketitle
	\begin{abstract}
		The long timescale associated with ligand residence times renders their computation challenging. Therefore, the influence of factors like solvation and steric hindrance on residence times are not fully understood. Here, we demonstrate in a set of model host-guest systems that the recently developed Gaussian Mixture Based Enhanced Sampling allows residence times to be computed and enables understanding their unbinding mechanism. We observe that guest unbinding often proceeds via a series of intermediate states that can be labelled by the number of water molecules present in the binding cavity. And in several cases the residence time is correlated to the water trapping times in the cavity. 
	\end{abstract}
\end{@twocolumnfalse}
]


\section*{Introduction}
Having a measure of the pharmacological efficacy of a molecule is crucial to a successful drug design process. The standard approach is to use binding affinities to estimate drug efficacy. 
However, recent studies indicate that kinetics and in particular drug unbinding rates 
could be more closely correlated to drug efficacy. Therefore, effort has been put in understanding the relationship between a drug and its unbinding kinetics \cite{Ladbury1996,Tiwary2015,Schuetz2017,Tang2017,Bruce2018,Kokh2018,Miao2020}. Although it is possible to measure thermodynamic and kinetic properties of drug binding in the laboratory, such studies are expensive and time consuming. It is therefore important to be able to compute residence times and relate them to the microscopic details of the unbinding process\cite{Jorgensen2004,DeVivo2016,Durrant2011}. 
Unfortunately, drug unbinding processes often occur on timescales that are difficult to investigate using conventional molecular dynamics. Thus enhanced sampling methods are frequently used. 
A class of these enhanced sampling methods involves the addition of an external bias potential acting on a few appropriately chosen degrees of freedom often referred to as Collective Variables (CVs)\cite{Torrie1977,Laio2002,GMM-US,adaptiveUS,Valsson2014,RAVE,OPES}. If the CVs  are well chosen, such methods\cite{SGOOP,Sultan2018,HLDA,Bonati2020}
can greatly facilitate the calculation of binding affinities and other thermodynamic properties. 

However, despite recent advancements in enhanced sampling methodologies, dynamic properties cannot be straightforwardly computed as the bias that is added to enhance sampling alters the natural dynamics of the system. Sometime ago \citeauthor{conformationalFlooding}\cite{conformationalFlooding} and \citeauthor{hyperdynamics}\cite{hyperdynamics} suggested that obtaining reaction rates between different metastable state was still possible if the bias added to the system did not act on the transition region. More recently, these ideas have been used in the context of enhanced sampling simulations and different strategies have been put forward to this effect\cite{infreqMetad,VarFlooding}.  However, imposing the condition that the transition state is not tainted by the bias is not simple. In order to alleviate this difficulty, we have recently proposed a new enhanced sampling method called Gaussian mixture based enhanced sampling (GAMBES)\cite{GAMBES}, in which the bias can be made null in the transition state region in a rather simple way. In this paper, we would like to demonstrate how GAMBES could be used in computing drug unbinding rates ($k_{off}$) in a set of 12 realistic host-guest systems that have been part of the SAMPL5 challenge\cite{Yin2016,Sullivan2016} and whose static properties were also studied in our group\cite{Rizzi2021}. In particular, in our previous work the focus was on the role of water. In line with other studies it was found that water solvation, both of the binding cavity and of the ligand, plays a very important role in binding. We will address also here the role of water, however this time we shall focus on the effect of water on dynamical rather than thermodynamical properties. In so doing, we shall profit from the tools that were set up in Ref. \citenum{Rizzi2021} to describe water solvation. Using GAMBES we are able to calculate not only the $k_{off}$ but also obtain a detailed description of the ligand unbinding dynamics. 

\section*{Method}
The physical picture in which the use of GAMBES is appropriate is one of a physical system that spends most of its time in a finite number of metastable states and makes only rare transitions from one state to another. In other words, unbiased trajectories started from any of the metastable states remain trapped there for a long time. In GAMBES, one takes advantage of this behaviour and estimates the local probability densities by fitting a Gaussian Mixture Model (GMM) to the data collected from these localized trajectories. In principle, these GMMs could be fitted to the full configurational space ${\boldsymbol{R}}$. However, such a  bias is ineffective in driving transitions out of the local island of stability because of its high dimensionality. In order to improve the bias efficiency, $N_d$ set of descriptors ${\boldsymbol{d}({\boldsymbol{R}})} \equiv \{ d_i({\boldsymbol{R}}); i \in \{ 1, N_d \}\}$ are introduced to reduce the dimensionality of the bias. However, care must be taken that all the metastable states of interest can still be distinguished by their $\boldsymbol{d}$ values. Furthermore, a wise choice of $\boldsymbol{d}$ enforces the original symmetries of the system like invariance to translations, rotations and atomic permutations.

Therefore, the data obtained from an unbiased simulation of a given metastable state $i$ is used to fit a GMM and estimate its probability distribution as: \\ 
\begin{equation}
    p^i({\boldsymbol{d}}({\boldsymbol{R}})) = \sum_{k=1}^{K}{\pi_k^i \mathcal{N}({\boldsymbol{d}}|\mu_k^i,\Sigma_k^i)}
    \label{gmm_state}
\end{equation}
where $K$ is the number of multivariate Gaussians required to describe the probability density of state $i$ and $\pi^k$ are the weights attributed to each of these Gaussians such that $\sum_k^{K} \pi^k=1$. The parameter $K$ is obtained using a Bayesian Inference Criterion\cite{schwarz1978}. The global ${\boldsymbol{d}}$ probability density of the system, $P(\boldsymbol{d}({\boldsymbol{R}}))$, can then be approximated as a weighted sum of the local probability densities, $p^i(\boldsymbol{d}({\boldsymbol{R}}))$s. Thus, \\
\begin{equation}
    P({\boldsymbol{d}}({\boldsymbol{R}})) \approx \sum_{i=1}^M \frac{1}{Z^i} p^i({\boldsymbol{d}}({\boldsymbol{R}}))
\end{equation}
where $Z^i = \int d{\boldsymbol{d}({\boldsymbol{R}})}\ p^i({\boldsymbol{d}}({\boldsymbol{R}}))$ and $M$ denotes the total number of metastable islands.
The bias is computed from this model probability density using the relation, \\
\begin{equation}
   V({\boldsymbol{d}}({\boldsymbol{R}}))=\frac{1}{\beta} \log{P({\boldsymbol{d}}({\boldsymbol{R}}))} 
\end{equation}
where $\beta=1/k_BT$ is the inverse temperature. 
After making a simple transformation and dropping the irrelevant constant $-\log{(Z_1)}/\beta $, the bias can be rewritten as, \\
\begin{equation}
    V({\boldsymbol{d}}({\boldsymbol{R}}))=\frac{1}{\beta} \log{\sum_{i=1}^M \frac{Z^1}{Z^i} p^i({\boldsymbol{d}}({\boldsymbol{R}}))}
\end{equation}
This factor $\frac{Z^1}{Z^i}$ gives an estimate of the relative occupation of the metastable states. It is not necessary to know this factor a priori as it can be calculated self consistently using, \\
\begin{equation}
    \frac{Z^1}{Z^i} = \frac{\sum_t \log(p^1({\boldsymbol{d}})) \ e^{ \beta ( V({\boldsymbol{d}}) )}}{\sum_t  \log(p^i({\boldsymbol{d}})) \ e^{ \beta ( V({\boldsymbol{d}}) )}}
\end{equation}

The value of the bias thus constructed is ill defined in the regions between the metastable states due to negligible overlap between the local probability densities and the exponential tails of the Gaussian distributions. To remedy this problem, we cutoff the Gaussian tails to a preassigned value $p_c$. This in turn introduces an upper value to the added bias ($V_c = 1/\beta \ log(p_c)$). The cutoff $V_c$ has a role in a way similar to that played by the free energy cutoff in variational flooding\cite{VarFlooding}. It is thanks to the introduction of this cutoff that the GAMBES bias goes to zero in regions outside the metastable states.  We can then obtain the first passage times using the rescaling proposed in Refs. \citenum{conformationalFlooding,hyperdynamics},
\begin{equation}
    \tau_r = \tau_{MD} \big< e^{\beta V({\boldsymbol{d}})} \big>_V
    \label{eq:fpt}
\end{equation}
where $\tau_{MD}$ is the apparent residence time measured during the biased simulation while $\tau_r$ is its corresponding physical value. 
\begin{figure}
    \centering
    \includegraphics[width=0.8\linewidth]{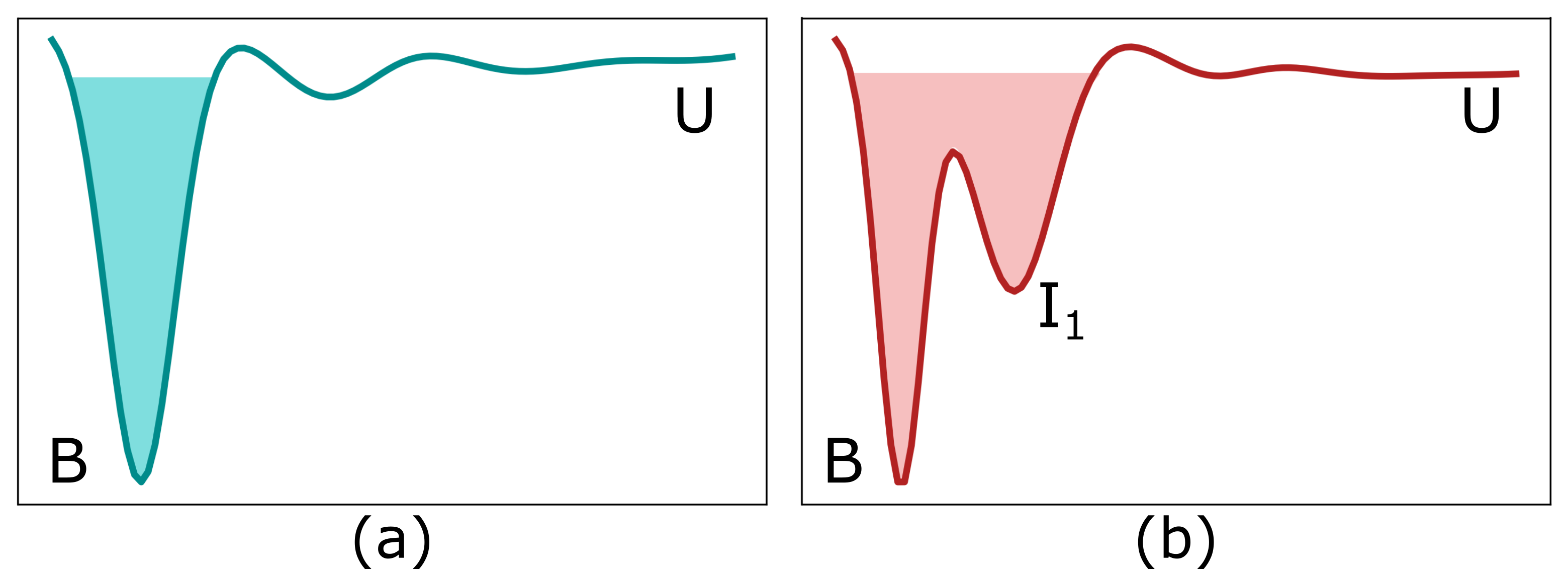}
    \caption{Schematic representation of free energy landscapes for drug binding-unbinding studies. Shaded region in the plots represent states that have been filled by the bias. (a) no metastable intermediates present, (b) metastable intermediate present.}
    \label{fig:fes_b_u}
\end{figure}

The transition between the bound and unbound states, B and U, occurs in one of the two ways shown in Figure \ref{fig:fes_b_u}. In the case of a scenario like that of Figure \ref{fig:fes_b_u}a, the system makes a direct transition from B to U. In such a case the rate calculation proceeds as described in Ref. \citenum{GAMBES}. That is, first we fit a GMM to the $N_d$ dimensional data collected from an unbiased trajectory started at B. We then run multiple simulations starting from state B with the bias constructed using an appropriately chosen cutoff. The residence times thus obtained are fitted to a Poisson distribution to obtain the rate.

However, if we are in a more complex scenario like the one schematically illustrated in Figure \ref{fig:fes_b_u}b,
an intermediate $I_1$ is present such that the barrier for going from $B$ to $I_1$ is large and so is the one from $I_1$  to $U$. Thus, if we bias only B the system will go to $I_1$ and remain trapped there as it feels no bias when in $I_1$ and it is surrounded by large barriers. Therefore, the trajectory generated while the system is trapped in $I_1$ can be used to model the $I_1$ probability distribution. Using the GAMBES machinery, a new bias that encompasses states B and $I_1$ can be created as indicated by the shaded region in Figure \ref{fig:fes_b_u}b. This combined bias can then be used to drive the system into the unbound state and the $k_{off}$ can finally be computed. 
In general, the number of intermediate states can be larger than one. In such a case, the procedure has to be continued until all the intermediate states are explored and the ligand exits the binding site. 

\section*{Results and discussion}
\begin{figure}
    \centering
    \includegraphics[width=0.95\linewidth]{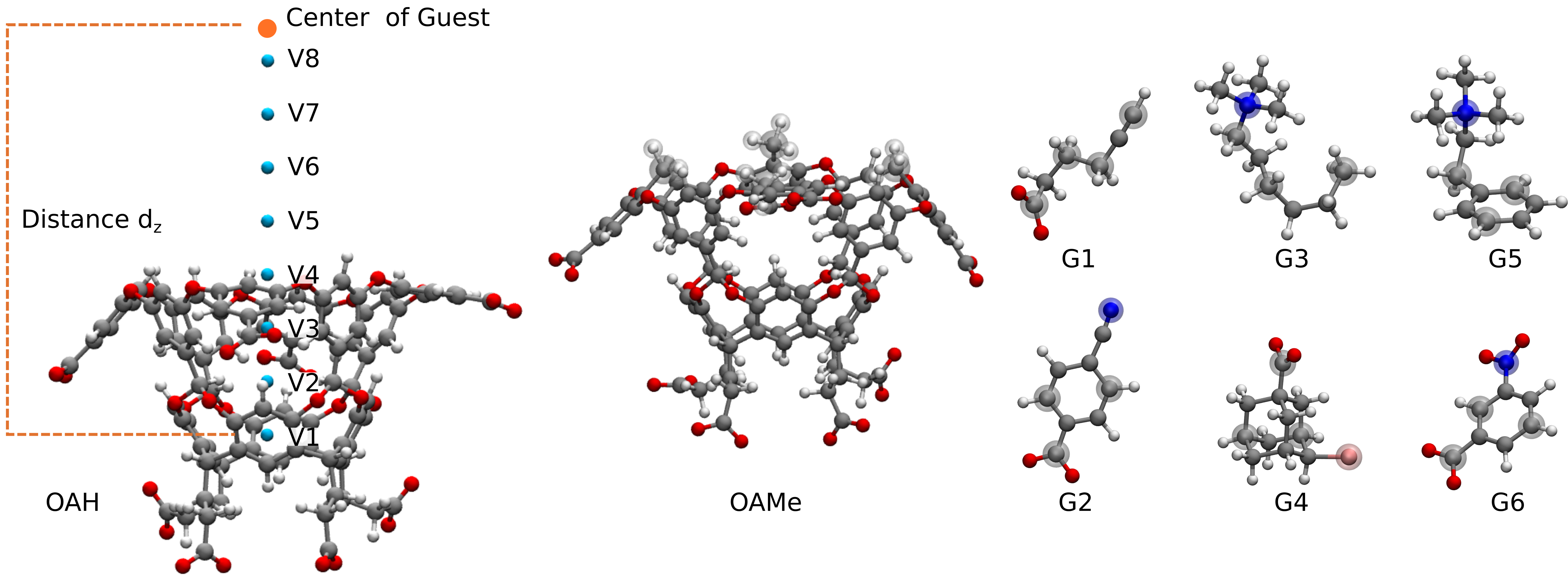}
    \caption{ The water solvation of the host and guest were characterized using coordination of Oxygen atoms of all the water molecules in the box with 12 atoms: 8 virtual points (shown as blue balls) and 4 guest atoms (highlighted ones). The z-component of the distance ($d_z$) between the center of the guest molecule and the virtual atom V1 was used to measure the position of the guest w.r.t the binding site in the host. These 13 descriptors, the 12 coordinations and the $d_z$ distance were then used to construct GAMBES bias.}
    \label{fig:descriptors}
\end{figure}

\begin{figure}[htbp]
    \centering
    \includegraphics[width=0.95\linewidth]{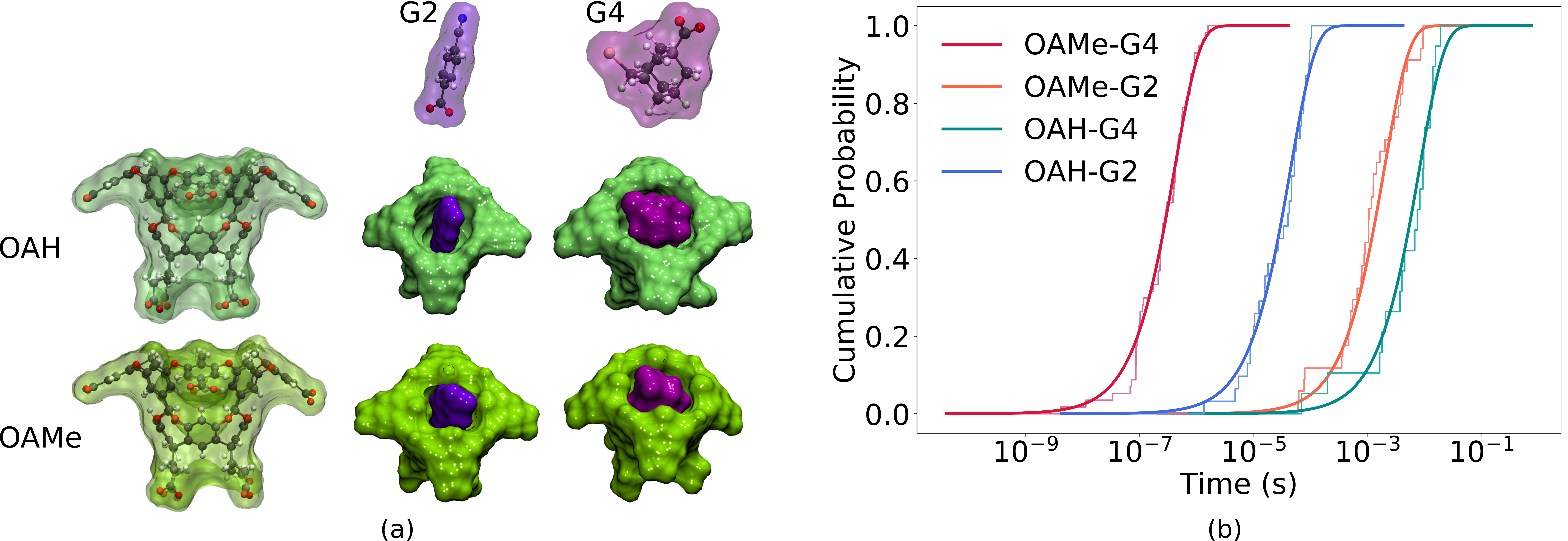}
    \caption{(a) Surface plots of the two guests (``thin'' G2 and ``fat'' G4) inside the two hosts (``narrow'' OAMe and ``wide'' OAH) showing space available in the binding pose. (b) Unbinding rates}
    \label{fig:hostguestsurfrate}
\end{figure}

We study all 12 Octa-acid host-guest systems of the SAMPL5 challenge. The interactions in these systems resemble those present in more practical drug binding scenarios. Hence, they are considered prototypical systems that help in understanding some of the features of drug binding. These host-guest systems consist of the combination of two different hosts (see Figure \ref{fig:descriptors}): the ``wide'' mouthed Octa-acid calixarene host (OAH) and the ``narrow'' mouthed methyl substituted Octa-acid (OAMe) also known as tetra-endo methyl octa-acid (TEMOA) with the 6 guests. Although we studied all the 6 guest molecules that are part of the SAMPL5 challenge, in the main text we discuss only two of these guests (see Figure \ref{fig:hostguestsurfrate}): a ``thin'' molecule G2 and a ``fat'' one G4 in combination with the two hosts. The resulting 4 systems are representative of the different scenarios that are to be encountered in the full set of the 12 examples. Results for all the other 8 systems can be found in the Supporting Information.

As discussed earlier, we build on the study of \citeauthor{Rizzi2021}\cite{Rizzi2021} where the water solvation in the binding pocket and the solvation of the guest molecules were discussed in detail. In their work, \citeauthor{Rizzi2021} used 12 descriptors (see Figure \ref{fig:descriptors} ) to characterize water solvation in the bound state and the unbound state to train a neural network based Deep-LDA model\cite{Bonati2020} and obtained a CV that describes the solvation effects via the non-linear dependence of the CV on the descriptors.
Then, they biased this Deep-LDA CV and the vertical distance ($d_z$) between the binding pocket and the center of mass of the guest molecule using the On-the-fly Probability Enhanced Sampling (OPES) method\cite{OPES} to obtain free energy surfaces and calculate binding affinities. 
In order to calculate the unbinding rate for these host-guest systems, we start by modelling the bound state probability density using a GMM. As discussed earlier, the GMMs should be fitted in the reduced dimensional descriptor space rather than the full $\boldsymbol{R}$ space. 
In principle, we could use the Deep-LDA CV\cite{Rizzi2021} and the vertical distance $d_z$ to form a GAMBES bias. In scenarios like that of Figure \ref{fig:fes_b_u}a this works rather well. However, when one starts with systems like that of Figure \ref{fig:fes_b_u}b, one encounters difficulties like those described in Ref. \citenum{GAMBES}. While the remedies suggested in Ref. \citenum{GAMBES} would solve this problem, we found more efficient and illuminating to express the GAMBES bias as a function of all the 12 solvation descriptors and $d_z$. 
In such a way the intermediates can be clearly distinguished and the unbinding process made more easily interpretable. 
In order to calculate the drug affinity a finer resolution of the bound state into substates is not necessary. 

In order to evaluate the unbinding rate, we followed the approach outlined earlier, i.e., fitted GMM to all the intermediate states discovered during the unbinding process and used multi-state static biases to drive the guests from their most stable bound states to the unbound states. 
Following the procedure described earlier, we then calculate the residence times. In Figure \ref{fig:hostguestsurfrate}b, the plots of the cumulative distribution of the rescaled first passage times ($\tau_r$) are shown for the systems under consideration here, and in Table \ref{tab:alldata} we summarize all the parameters obtained by fitting the Poisson distribution.

\begin{table}[h]
	\centering
	\resizebox{\linewidth}{!}{%
		\begin{tabular}[width=1.9\linewidth]{|c|c|c|c|c|}
			\hline
			& \textbf{OAMe-G2} & \textbf{OAMe-G4} & \textbf{OAH-G2} & \textbf{OAH-G4} \\ 
			\hline
			\textbf{Mean of transition time $\boldsymbol{\mu}$ [s]} & ($2.18 \pm 0.45 $) x $10^{-3}$ & ($4.13 \pm 0.49$) x $10^{-7}$ & ($4.29 \pm 0.59 $) x $10^{-5}$ & ($7.77 \pm 1.27$) x $10^{-3}$ \\ 
			\hline
			\textbf{Std Dev of transition time $\boldsymbol{\sigma}$ [s]} & 2.60 x $10^{-3}$ & 3.68 x $10^{-7}$ & 3.27 x $10^{-5}$ & 5.55 x $10^{-3}$ \\
			\hline
			\textbf{Median of transition time $\boldsymbol{t_m}$ [s]} & 1.07 x $10^{-3}$ & 2.85 x $10^{-7}$ & 4.11 x $10^{-5}$ & 7.65 x $10^{-3}$ \\
			\hline
			$\boldsymbol{\tau}$ \textbf{[s]} & 2.02 x $10^{-3}$ & 4.22 x $10^{-7}$ & 4.57 x $10^{-5}$ & 8.35 x $10^{-3}$ \\
			\hline
			$\boldsymbol{k_{off} [s^{-1}]}$ & 4.95 x $10^{+2}$ & 2.37 x $10^{+6}$ & 2.19 x $10^{+4}$ & 1.20 x $10^{+2}$ \\
			\hline
			\textbf{p-value of KS statistics} & 0.74 & 0.9 & 0.88 & 0.51 \\
			\hline
			$\boldsymbol{\Delta F^{exp}}$ \textbf{[kcal/mol]} \cite{Yin2016}& $-5.24 \pm 0.05$ & $-2.38 \pm 0.02$ & $ -4.25 \pm 0.01$ & $-9.37 \pm 0.00$  \\
			\hline
		\end{tabular}
	}
	\caption{Unbinding rates and parameters obtained upon fitting Poisson distribution to the cumulative probability distribution of the rescaled first passage times}
	\label{tab:alldata}
\end{table}

\begin{figure}[htbp]
    \centering
    \includegraphics[width=0.9\linewidth]{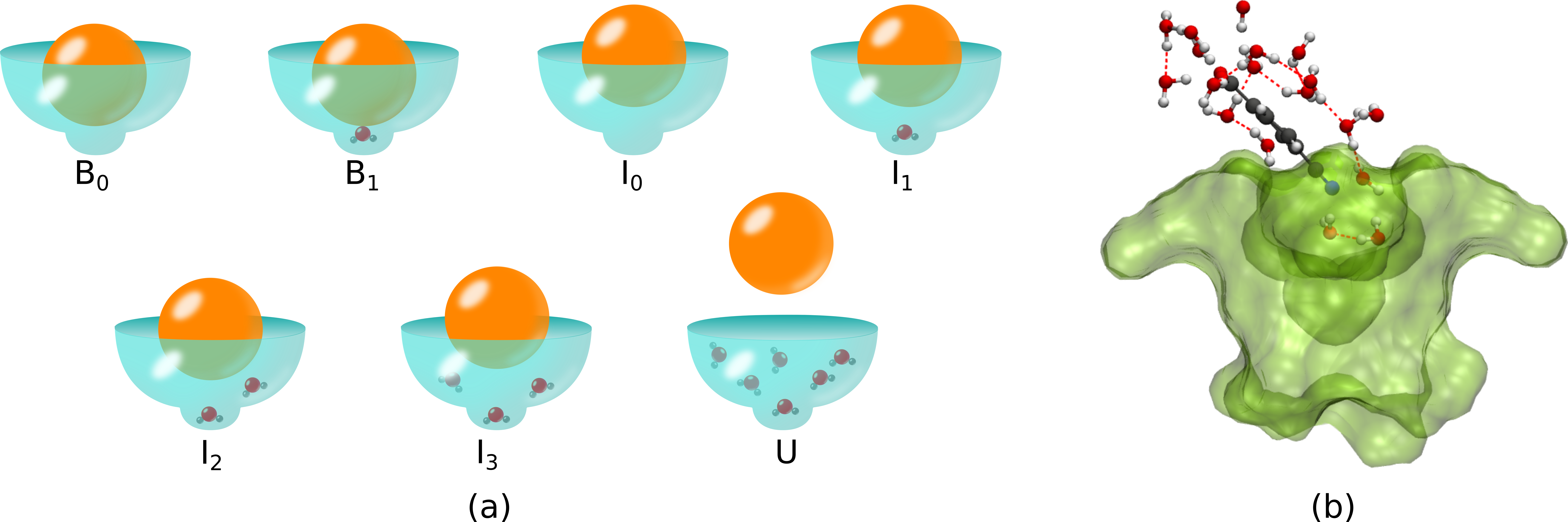}
    \caption{(a) Schematic representation of the different types of metastable states. (b) Snapshot of a chain of water forming around guest G2 in host OAMe.}
    \label{fig:states}
\end{figure}

In Figure \ref{fig:states}a, we have summarised all possible intermediate states that are visited during the unbinding process. 
Quite surprisingly, despite the dissimilarities in the host and guest interactions, the intermediate states discovered had many features in common.
All states can be classified on whether they are bound (B), intermediate (I) or unbound (U). There could be multiple B and I states. In all cases, they can be distinguished by their number of water molecules trapped in the cavity. Such number will be used below as a subscript. A state is considered bound if $d_z$ is close to its experimental value. The bound states thus classified have low and competing free energies. A molecule is considered unbound if  $d_z$ is greater than $1.1 \ \text{\AA}$.
Although, these states were visited, their lifetime and relative heights varied greatly from system to system. 

The simplest unbinding mechanism was observed in the case of OAH-G2 that is a combination of the ``thin'' guest G2 and the ``wide'' host OAH (Figure \ref{fig:hostguestsurfrate}a). In this case, the volume accessible to water molecules is large even when the guest is bound and so water molecules can move in and out of the pocket easily. For this reason, states $B_0$, $B_1$, $I_1$, $I_2$, $I_3$ are separated by  low free energy barriers ($<k_BT$) and can easily interconvert. Therefore, host solvation did not play a critical role. We note a feature that is common to most host-guest systems, before the guest exits the pocket a chain of water molecules forms around the guest molecule and kind of scoops it outside the pocket (see Figure \ref{fig:states}b). 

In the case of OAMe-G2, the thin guest G2 is inserted in the narrower host OAMe (Figure \ref{fig:hostguestsurfrate}a). As a consequence of the more difficult access, the barrier between states $B_0$, $B_1$, $I_1$, $I_2$, $I_3$ are higher here than in the previous case and these states interconvert on a much longer time scale. This case was similar to that in Figure \ref{fig:fes_b_u}b, and although it was possible to move downhill from the intermediate states to $B_0$ in an unbiased simulation, depositing bias at each intermediate state was necessary to move uphill towards U. At this stage, the transition from one state to the other is well signalled by abrupt changes in the value of the descriptors. 

The case of OAH-G4 where the ``fat'' guest G4 is inside the ``wide'' host OAMe is similar to OAMe-G2 case. Even though the methyl groups are absent, the guest positioned at the entrance of the host still faces additional steric hindrance due to its larger size (Figure \ref{fig:hostguestsurfrate}a). In this case, states $B_0$, $B_1$, $I_1$, $I_2$ are very stable and it was not possible either to move uphill or downhill without depositing bias in these states. 

Finally in OAMe-G4 consisting of the ``fat'' guest G4 and the ``narrow'' host OAMe, not only does the methyl group constrain the entrance, but the guest also contributes to steric hindrance (Figure \ref{fig:hostguestsurfrate}a). Consequently, the guest molecule could be positioned either inside or partially solvated just outside the cavity. The free energy barrier between $B_0$ and $I_0$ was significantly higher than in all the cases. The barriers between 
the states $I_0$, $I_1$, $I_2$, $I_3$ are low ($< k_BT$) and they can be considered to be part of one single basin. Thus, the scenario of Figure \ref{fig:fes_b_u}a applies. At this point, it is important to state that the value of distance $d_z$ for the intermediates here were significantly higher than that for all the other cases and in these intermediate states the guest molecule was almost out of the pocket. At variance with the other cases, the significant step during the unbinding process was the transition from $B_0$ to $I_0$, and in this system, water did not seem to play an important role in unbinding in this system.

\section*{Conclusion}
From all the above observations we can conclude that in the two extremes of very little hindrance at the entrance of the pocket (OAH-G2) or very high hindrance (OAMe-G4), the solvation of the host pocket does not play any major role in unbinding. However, when the entrance to the pocket is sufficiently narrow, water molecules get trapped inside creating metastable intermediate states and thereby increasing the residence times of the guests in the pocket. In such cases, host solvation plays an important role in unbinding. In contrast to host solvation, guest solvation or more importantly, the formation of a chain of water molecules around the guest almost always precedes the sliding of the guest molecule out of the pocket. Similar phenomena has also been reported in recent literature\cite{Schiebel2018,Schmidtke2011,Englert2010,Ewell2008,Tiwary2017,Lukac2021}. Nonetheless, the exact role of water depends on the steric factors arising steric hindrance of the host or the guest itself. 
Although previous studies\cite{Tiwary2015,Makarov2000} revealed there is a relationship between residence times and steric constraints between guest molecule and host pocket, we find that this is not as straightforward. Residence time does increase with an increase in steric constraints between guest and host-entrance, but if these steric constraints are increased beyond a certain limit the binding pose is destabilized and residence time decreases.
We do not know of any other study calculating the rates or residence times in these systems, but the trends that we observe in our calculated residence times are correlated to those reported for binding affinities both using experimental and theoretical methods\cite{Yin2016,Rizzi2021}. 

\begin{acknowledgement}
The authors thank Valerio Rizzi for providing the input files and DeepLDA CVs for all the systems. JD thanks Valerio Rizzi, Narjes Ansari and Umberto Raucci for useful discussions and for carefully reading the manuscript. 
\end{acknowledgement}

\newpage
\onecolumn
\begin{center}\section{Supporting information}\end{center}
\subsection{Computational details}
The simulation parameters were the same as the ones used by \citeauthor{Rizzi2021} and the input files were taken from the public repository \url{https://github.com/michellab/Sire-SAMPL5}. Similar to their work, we used the GAFF\cite{Wang2004} forcefield with RESP\cite{Bayly1993} charges and TIP3P model\cite{Jorgensen1983} for water molecules. The timestep for running the molecular dynamics trajectories was 2 fs. The simulations were carried out in the NVT ensemble and the velocity rescaling thermostat\cite{Bussi2007} with time constant 0.1 ps was used to maintain a constant temperature of 300K. A cubic simulation box of length 40 $\AA$ was constructed, with the host and the guest molecules solvated in 2100 water molecules. Sodium ions were included in the box to enforce neutrality. All the simulations were carried out using GROMACS 2020.4\cite{GROMACS} and PLUMED 2.7\cite{Plumed2}. The PLUMED code was compiled with Pytorch version 1.4 and the GAMBES code that is available in the public repository \url{https://github.com/Jayashrita/GAMBES}. The trajectories were all visualized using VMD\cite{VMD}.
\subsection{Rates of all systems}
\begin{figure}[h]
    \centering
    \includegraphics[width=0.5\columnwidth]{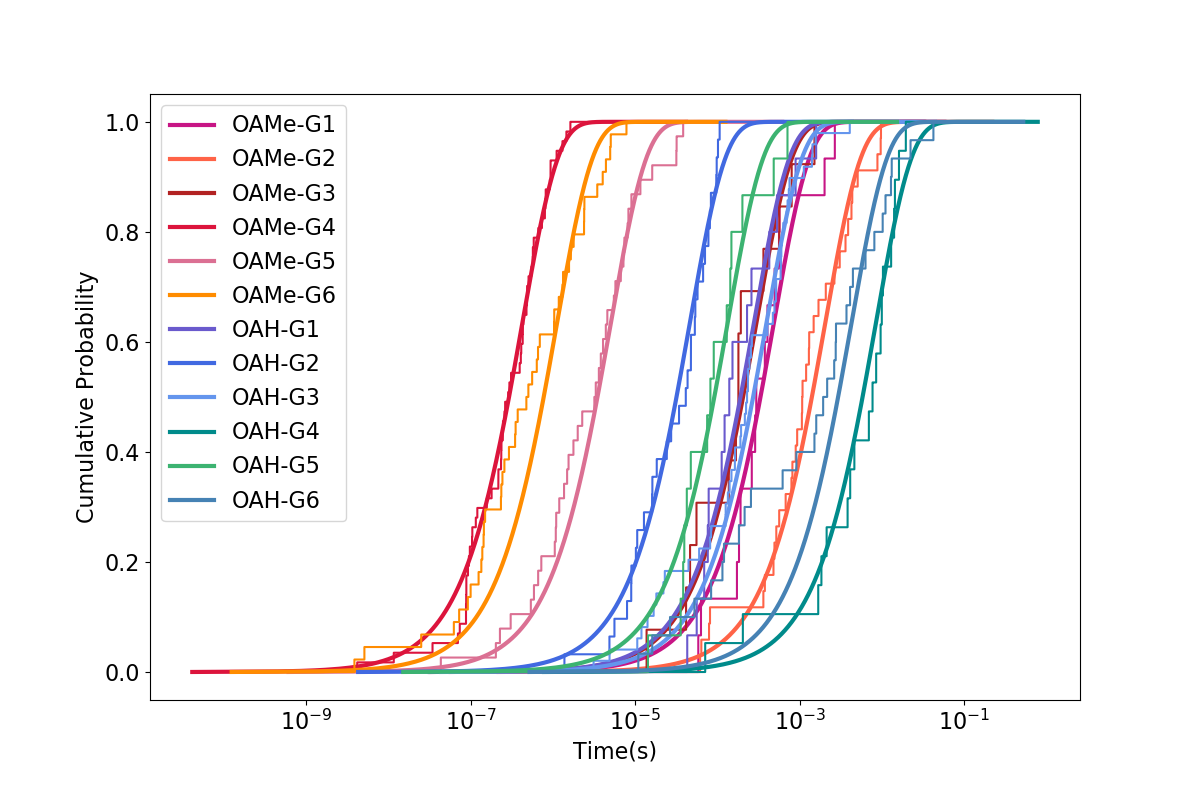}
    \caption{Poisson distribution fitted to the first passage times obtained for all the cases}
    \label{fig:rates1}
\end{figure}

The data for the rates extracted for all the systems have been summarized in Figure \ref{fig:rates1} and Table \ref{tab:alldata1}. 
In the following subsection, we describe briefly the mechanism of unbinding for the 8 systems not discussed in the main text.
\subsection{Mechanisms}

\subsubsection{Guest G1}
In OAMe-G1, like the ``thin'' guest G2, the guest here is "thin" too and is inserted in the ``narrow'' host OAMe where the entrance to the cavity is narrower. Even though the molecule is thinner than G2, it is flexible and some of its folded configurations occupy more volume like the "fat" guest G4. The mechanism in this case was more similar to OAMe-G2 where the barrier between states $B_0$, $B_1$, $I_1$, $I_2$, $I_3$ are higher here than in OAH-G2 but lower than OAMe-G2. 
In OAH-G1, the mechanism is quite similar to OAMe-G1 while the barrier separating the intermediates were a bit lower than in the OAMe-G1 case. Although the guest is thin like G2, it is not planar and so water can not freely move in and out of the host pocket in this case. 
\subsubsection{Guest G3}
G3 is flexible on one end like G1 but is bulky on the other end like G4. However, unlike G4, the bulky part here is almost always outside the cavity and solvated. The bulky part acts as an additional hindrance like the methyl groups of OAMe host. As a result water molecules cannot move in and out easily as in the case of G2, but the part of the guest that is inside the cavity does not face as much hindrance with the host as in G4. The unbinding of OAMe-G3 and OAH-G3 were very similar to OAMe-G2 case. \\

\subsubsection{Guest G5}
G5 is rigid and its one end is planar like G2 and inside the cavity while the other end is solvated and outside like G3.In OAMe-G5, because of the steric hindrance between the bulky end and the methyl groups, the unbinding mechanism is similar to OAMe-G4. However, as the bulky group is outside the cavity, water molecules can still enter the cavity to a small extent. Only the guest solvation and formation of water chain seemed to play a role in this unbinding. While in OAH-G5, the unbinding was more similar to OAMe-G2. The wider egress of the host allowed easier entry of water molecules, however the water molecules stayed trapped longer due to the bulky end of the guest molecule. 

\begin{table*}[h]
    \centering
    \resizebox{0.8\columnwidth}{!}{%
    \begin{tabular}[width=\textwidth]{|c|c|c|c|c|c|c|}
    \hline
     System & Mean $\mu$ [s] & Std Dev $\sigma$ [s] & Median $t_m$ [s] & $\tau$ [s] & $k_{off}$ [$s^{-1}$]& p-value \\
     \hline
        OAMe-G1 & ($5.71 \pm 1.85 $) x $10^{-3}$  & 7.16 x $10^{-4}$ & 1.85 x $10^{-4}$ & 4.80 x $10^{-4}$ & 2.09x $10^{+3}$ & 0.52 \\  \hline
        OAMe-G2 &  ($2.18 \pm 0.45 $) x $10^{-3}$  & 2.60 x $10^{-3}$ & 1.07 x $10^{-3}$ & 2.02 x $10^{-3}$ & 4.95x $10^{+2}$ & 0.74 \\  \hline
        OAMe-G3 & ($2.65 \pm 0.72$) x $10^{-4}$  & 3.57 x $10^{-4}$ & 1.41 x $10^{-4}$ & 2.33 x $10^{-4}$ & 4.29 x $10^{+3}$ & 0.8 \\  \hline
        OAMe-G4 & ($4.13 \pm 0.49$) x $10^{-7}$  & 3.68 x $10^{-7}$ & 2.85 x $10^{-7}$ & 4.22 x $10^{-7}$ & 2.37 x $10^{+6}$ & 0.9  \\  \hline
        OAMe-G5 & ($6.11 \pm 1.44$) x $10^{-6}$  & 8.88 x $10^{-6}$ & 3.25 x $10^{-6}$ & 5.01 x $10^{-6}$ & 2.0 x $10^{+5}$ & 0.94  \\  \hline
        OAMe-G6 & ($1.25 \pm 0.25$) x $10^{-6}$ & 1.69 x $10^{-6}$ & 4.89 x $10^{-7}$ & 1.11 x $10^{-6}$ & 8.99 x $10^{+5}$ &  0.74  \\  \hline
        OAH-G1 & ($3.17 \pm 1.04 $) x $10^{-4}$  & 4.03 x $10^{-4}$ & 1.39 x $10^{-4}$ & 2.74 x $10^{-4}$ & 3.65x $10^{+3}$ & 0.79 \\  \hline
        OAH-G2 & ($4.29 \pm 0.59 $) x $10^{-5}$  & 3.27 x $10^{-5}$ & 4.11 x $10^{-5}$ & 4.57 x $10^{-5}$ & 2.19 x $10^{+4}$ & 0.88   \\  \hline
        OAH-G3 & ($4.39 \pm 0.92$) x $10^{-4}$  & 6.43 x $10^{-4}$ & 2.28 x $10^{-4}$ & 3.9 x $10^{-4}$ & 2.56 x $10^{+3}$ & 0.87  \\  \hline
        OAH-G4 & ($7.77 \pm 1.27$) x $10^{-3}$  & 5.55 x $10^{-3}$ & 7.65 x $10^{-3}$ & 8.35 x $10^{-3}$ & 1.2 x $10^{+2}$ & 0.51  \\  \hline
        OAH-G5 & ($1.51 \pm 4.8$) x $10^{-4}$  & 1.85 x $10^{-4}$ & 0.81 x $10^{-4}$ & 1.32 x $10^{-4}$ & 7.59 x $10^{+3}$ & 0.6  \\  \hline
        OAH-G6 & ($4.28 \pm 1.2$) x $10^{-3}$  & 7.6 x $10^{-3}$ & 1.2 x $10^{-3}$ & 3.38 x $10^{-3}$ & 2.96 x $10^{+2}$ & 0.002  \\  \hline
    \end{tabular}
    }
    \caption{Unbinding rates and Poisson distribution parameters}
    \label{tab:alldata1}
\end{table*}
\subsubsection{Guest G6}
G6 too is rigid and not extremely bulky, but is bulkier than G2. Despite the thin nature of the guest, the positioning of its functional groups locks it inside the cavity and thus, does not allow water molecule to pass freely in and out, leading to the creation of trapped states. Due to this, the behaviour of this guest was very similar to G4 or G5. Thus, the unbinding of OAMe-G6 was like OAMe-G5 while the unbinding of OAH-G6 was like that in the case of OAH-G6.
\bibliography{GAMBES_rates}



\end{document}